\documentclass[aps,prl,twocolumn,showpacs,superscriptaddress]{revtex4}
\usepackage{graphicx}
\usepackage{dcolumn}
\usepackage{bm}
\usepackage{amsmath}
\usepackage{amssymb}

\begin{document}

\preprint{\vbox{ \hbox{Belle Preprint 2007-30}
                 \hbox{KEK   Preprint 2007-22}
                              }}

\title{\quad\\
[0.5cm] Search for $B \to h^{(*)} \nu \overline{\nu}$ Decays at Belle}

\begin{abstract}
We present a search for the rare decays $B \to h^{(*)} \nu \overline{\nu}$, where $h^{(*)}$ stands for a light meson.
A data sample of 535 million $B\overline{B}$ pairs 
collected with the Belle
detector at the KEKB $e^+e^-$ collider is used. 
Signal candidates are required to have an accompanying $B$ meson fully reconstructed in 
a hadronic mode 
and signal-side particles consistent with a single $h^{(*)}$ meson. 
No significant signal is observed and we set upper limits on the branching fractions
at 90\% confidence level.
The limits on $B^0 \to K^{*0}\nu\overline{\nu}$ and $B^+ \to K^+\nu\overline{\nu}$ decays
are more stringent than the previous constraints, while 
the first searches for
$B^0 \to K^0 \nu\overline{\nu}$, $\pi^0\nu\overline{\nu}$, $\rho^0\nu\overline{\nu}$, $\phi\nu\overline{\nu}$ and
$B^+ \to K^{*+}\nu\overline{\nu}$, $\rho^+\nu\overline{\nu}$ are reported.

\end{abstract}

\affiliation{Budker Institute of Nuclear Physics, Novosibirsk}
\affiliation{Chiba University, Chiba}
\affiliation{University of Cincinnati, Cincinnati, Ohio 45221}
\affiliation{Department of Physics, Fu Jen Catholic University, Taipei}
\affiliation{The Graduate University for Advanced Studies, Hayama}
\affiliation{Hanyang University, Seoul}
\affiliation{University of Hawaii, Honolulu, Hawaii 96822}
\affiliation{High Energy Accelerator Research Organization (KEK), Tsukuba}
\affiliation{Hiroshima Institute of Technology, Hiroshima}
\affiliation{University of Illinois at Urbana-Champaign, Urbana, Illinois 61801}
\affiliation{Institute of High Energy Physics, Chinese Academy of Sciences, Beijing}
\affiliation{Institute of High Energy Physics, Vienna}
\affiliation{Institute of High Energy Physics, Protvino}
\affiliation{Institute for Theoretical and Experimental Physics, Moscow}
\affiliation{J. Stefan Institute, Ljubljana}
\affiliation{Kanagawa University, Yokohama}
\affiliation{Korea University, Seoul}
\affiliation{Kyungpook National University, Taegu}
\affiliation{Swiss Federal Institute of Technology of Lausanne, EPFL, Lausanne}
\affiliation{University of Ljubljana, Ljubljana}
\affiliation{University of Maribor, Maribor}
\affiliation{University of Melbourne, School of Physics, Victoria 3010}
\affiliation{Nagoya University, Nagoya}
\affiliation{Nara Women's University, Nara}
\affiliation{National Central University, Chung-li}
\affiliation{National United University, Miao Li}
\affiliation{Department of Physics, National Taiwan University, Taipei}
\affiliation{H. Niewodniczanski Institute of Nuclear Physics, Krakow}
\affiliation{Nippon Dental University, Niigata}
\affiliation{Niigata University, Niigata}
\affiliation{University of Nova Gorica, Nova Gorica}
\affiliation{Osaka City University, Osaka}
\affiliation{Osaka University, Osaka}
\affiliation{Panjab University, Chandigarh}
\affiliation{Princeton University, Princeton, New Jersey 08544}
\affiliation{RIKEN BNL Research Center, Upton, New York 11973}
\affiliation{University of Science and Technology of China, Hefei}
\affiliation{Seoul National University, Seoul}
\affiliation{Sungkyunkwan University, Suwon}
\affiliation{University of Sydney, Sydney, New South Wales}
\affiliation{Tata Institute of Fundamental Research, Mumbai}
\affiliation{Toho University, Funabashi}
\affiliation{Tohoku Gakuin University, Tagajo}
\affiliation{Tohoku University, Sendai}
\affiliation{Department of Physics, University of Tokyo, Tokyo}
\affiliation{Tokyo Institute of Technology, Tokyo}
\affiliation{Tokyo Metropolitan University, Tokyo}
\affiliation{Virginia Polytechnic Institute and State University, Blacksburg, Virginia 24061}
\affiliation{Yonsei University, Seoul}

  \author{K.-F.~Chen}\affiliation{Department of Physics, National Taiwan University, Taipei} 
  \author{I.~Adachi}\affiliation{High Energy Accelerator Research Organization (KEK), Tsukuba} 
  \author{H.~Aihara}\affiliation{Department of Physics, University of Tokyo, Tokyo} 
  \author{V.~Aulchenko}\affiliation{Budker Institute of Nuclear Physics, Novosibirsk} 
  \author{T.~Aushev}\affiliation{Swiss Federal Institute of Technology of Lausanne, EPFL, Lausanne}\affiliation{Institute for Theoretical and Experimental Physics, Moscow} 
  \author{S.~Bahinipati}\affiliation{University of Cincinnati, Cincinnati, Ohio 45221} 
  \author{A.~M.~Bakich}\affiliation{University of Sydney, Sydney, New South Wales} 
  \author{V.~Balagura}\affiliation{Institute for Theoretical and Experimental Physics, Moscow} 
  \author{E.~Barberio}\affiliation{University of Melbourne, School of Physics, Victoria 3010} 
  \author{A.~Bay}\affiliation{Swiss Federal Institute of Technology of Lausanne, EPFL, Lausanne} 
% ...
\collaboration{The Belle Collaboration}


\maketitle
    
\tighten    



\begin{figure}[htpb]
\begin{center}
\includegraphics[width=4.2cm]{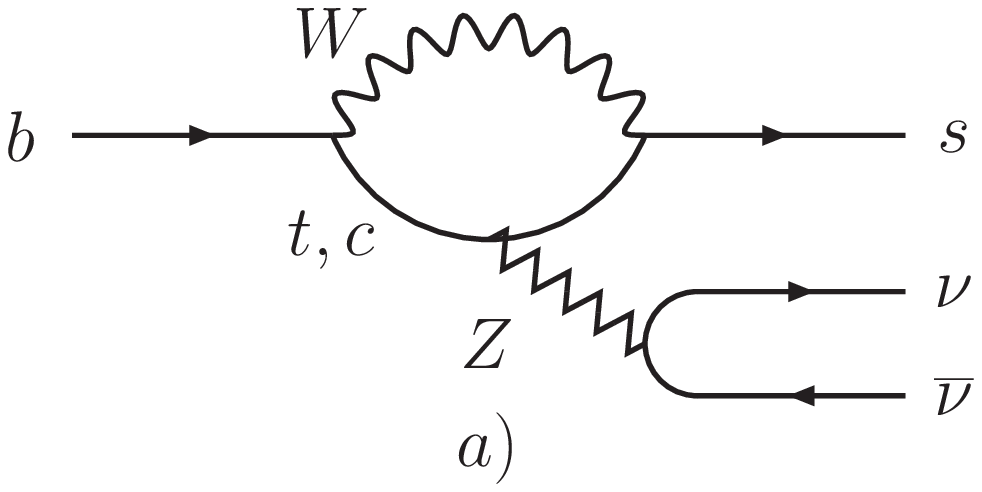} 
\includegraphics[width=4.2cm]{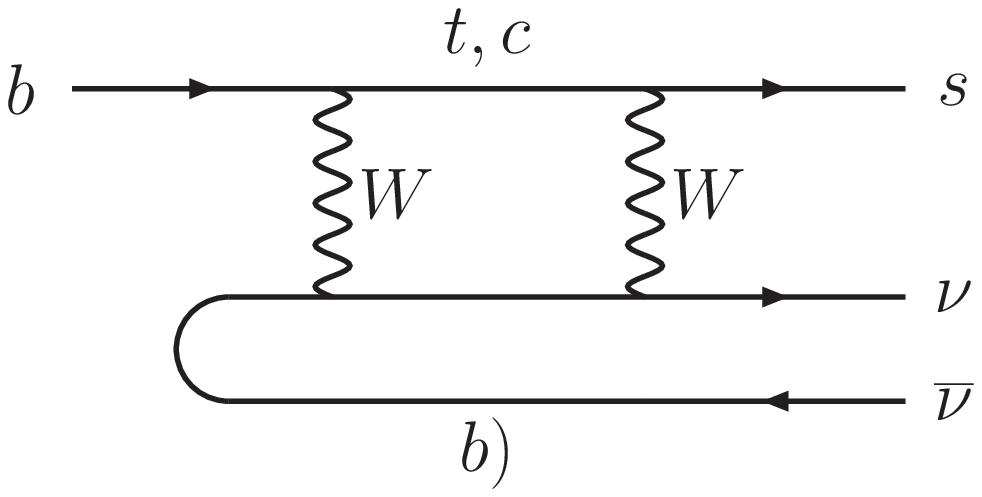}
\end{center}
\caption{The quark-level diagrams for $B \to K^{*} \protect\nu \overline{\protect\nu}$
decays.}
\label{fig:diagram}
\end{figure}

The decays $B \to K^{(*)} \nu \overline{\nu}$ 
proceed through the flavor-changing neutral-current process $b \to s \nu \overline{\nu}$, which is
sensitive to physics beyond the Standard Model (SM). 
The dominant SM diagrams are shown in Fig.~\ref{fig:diagram}. 
Similarly, the decays $B \to (\pi, \rho) \nu \overline{\nu}$ proceed through
$b \to d \nu \overline{\nu}$ processes. 
The SM branching fractions are estimated to be 
$1.3 \times 10^{-5}$ and $4 \times 10^{-6}$ for $B \to K^* \nu \overline{\nu}$
and $B \to K \nu \overline{\nu}$ decays~\cite{ref:buchalla}, respectively, 
and are expected to be much lower for other modes.
Theoretical calculation of the decay amplitudes for $B \to h^{(*)} \nu \overline{\nu}$
is particularly reliable, owing to the absence of long-distance
interactions that affect charged-lepton channels $B \to h^{(*)} l^+l^-$.
New physics such 
as SUSY particles or a possible fourth generation could
potentially contribute to the penguin loop or box diagram 
and enhance the branching fractions~\cite{ref:buchalla}.
Reference~\cite{ref:darkmatter} also discusses the possibility of
discovering light dark matter in $b\to s$ transitions with large
missing momentum.

Experimental measurements~\cite{ref:kll} of the $b\to s$ transitions with
two charged leptons are in good agreement with SM calculations~\cite{ref:buchalla}. Further
investigation of the forward-backward asymmetry in 
$B\to K^* l^+l^-$~\cite{ref:afb} is consistent with the SM
although the statistics are still limited. 
Due to the challenge of cleanly detecting rare modes with two final-state
neutrinos, only a few studies of $K^{(*)}\nu \overline{\nu}$ have been
carried out to date~\cite{Adam:1996ts,Browder:2000qr,Aubert:2004ws};
there is only one examination of the corresponding $b \to d$ transitions~\cite{Aubert:2004ws}. 
In this paper, we report our first search for the decays $B \to h^{(*)} \nu 
\overline{\nu}$ ($h^{(*)}$ stands for $K^+$, $K_S^0$, $K^{*0}$, $K^{*+}$, $\pi^+$, $\pi^0$, $\rho^0$, $\rho^+$, and $\phi$) 
using a 492 fb$^{-1}$ data sample recorded at the $\Upsilon(4S)$ resonance, 
corresponding to $535\times 10^6$ $B$-meson pairs. Charge-conjugate decays are implied throughout this
paper.

The Belle detector is a large-solid-angle magnetic spectrometer located at
the KEKB collider~\cite{ref:KEKB}, and consists of a silicon vertex detector
(SVD), a 50-layer central drift chamber (CDC), an array of 
aerogel threshold Cherenkov counters (ACC), a barrel-like arrangement of
time-of-flight scintillation counters (TOF), and an electromagnetic
calorimeter comprised of CsI(Tl) crystals (ECL) located inside a
superconducting solenoid that provides a 1.5~T magnetic field. An iron
flux-return located outside the coil is instrumented to detect $K_L^0$
mesons and to identify muons (KLM). The detector is described in detail
elsewhere~\cite{ref:belle_detector}.

Candidate $e^+e^- \to \Upsilon(4S) \to B\overline{B}$ 
events are characterized by a fully-reconstructed tag-side $B$ 
meson ($B_{\rm tag}$).
The remaining particles are assumed to be products of  
the signal-side $B$ meson ($B_{\mathrm{sig}}$). The $B_{\mathrm{tag}}$
candidates are reconstructed in one of the following modes: $B^0 \to
D^{(*)-} \pi^+$, $D^{(*)-}\rho^+$, $D^{(*)-}a_1^+$, $D^{(*)-}D_s^{(*)+}$, 
$B^+ \to \overline{D}{}^{(*)0} \pi^+$, $\overline{D}{}^{(*)0} \rho^+$,
$\overline{D}{}^{(*)0} a_1^+$, and $\overline{D}{}^{(*)0} D_s^{(*)+}$.
The $D^-$ mesons are reconstructed as $D^- \to K^0_S\pi^-$, $K_S^0\pi^-\pi^0$, 
$K_S^0\pi^-\pi^+\pi^-$, $K^+\pi^-\pi^-$, and $K^+\pi^-\pi^-\pi^0$. 
The following decay channels are included for $\overline{D}{}^0$ mesons: 
$\overline{D}{}^0 \to K^{+}\pi^{-}$, $K^+\pi^-\pi^0$, $K^+\pi^-\pi^+\pi^-$, 
$K_S^0\pi^0$, $K_S^0\pi^-\pi^+$, $K_S^0\pi^-\pi^+\pi^0$ and $K^-K^+$. 
The $D^{*-}$ ($\overline{D}{}^{*0}$) mesons are reconstructed as $\overline{D}{}^0 \pi^-$ 
($\overline{D}{}^0 \pi^0$ and $\overline{D}{}^0 \gamma$).
Furthermore, $D_s^{*+} \to D_s^+ \gamma$, $D_s^+\to K_S^0K^+$ and $K^+K^-\pi^+$ decays
are reconstructed. 
$B_{\mathrm{tag}}$ candidates are selected using the
beam-energy constrained mass $M_{\mathrm{bc}} \equiv \sqrt{E_{\mathrm{beam}}^{2} - p_{B}^{2}}$ 
and the energy difference $\Delta E \equiv E_{B} - E_{\mathrm{beam}}$, 
where $E_{B}$ and $p_{B}$ are the reconstructed energy and momentum
of the $B_{\mathrm{tag}}$ candidate in the $\Upsilon(4S)$ center-of-mass
(CM) frame, and $E_{\mathrm{beam}}$ is the beam energy in this frame.

We require $B_{\mathrm{tag}}$ candidates satisfy
the requirements
$M_{\mathrm{bc}}>5.27$~GeV/$c^2$ and $-80$~MeV~$<\Delta E< 60$~MeV. If there are multiple 
$B_{\mathrm{tag}}$ candidates in an event, the candidate with the smallest 
$\chi^{2}$ based on the deviations from the nominal values of $\Delta E$, the 
$D$ meson mass, and the mass difference between the  $D^{*}$ and the $D$ (for 
candidates with a $D^*$ in the final state) is chosen. 
We reconstruct $7.88 \times 10^5$
and $4.91 \times 10^5$ charged and
neutral $B$ mesons, respectively.


The particles in the event not associated with the $B_{\rm tag}$ meson 
are used to reconstruct a $B_{\rm sig} \to h^{(*)}\nu\overline{\nu}$ candidate.
Prompt charged tracks are required to have a 
maximum distance to the interaction point (IP) of 5 cm in the beam direction ($z$), 
of 2 cm in the transverse plane ($r$--$\phi $), and a minimum momentum of 0.1 GeV/$c$ in the transverse plane.
We reconstruct $K^\pm$ ($\pi^\pm$)
candidates from charged tracks
having a kaon likelihood greater than 0.6 (less than 0.4) with an efficiency of 84--91\% (87--92\%).
The kaon likelihood is defined by $\mathcal{R}_{K}\equiv \mathcal{L}_{K}/(\mathcal{L}_{K}+\mathcal{L}_{\pi })$,
where $\mathcal{L}_{K}$ ($\mathcal{L}_{\pi }$) denotes a combined likelihood
measurement from the ACC, the TOF, and a $dE/dx$ from the CDC for the $K^{\pm }$ ($\pi ^{\pm }$) tracks.
Pairs of oppositely charged tracks are used
to reconstruct $K^0_S \to \pi^+\pi^-$ decays, 
with an invariant mass that is within $\pm$15 MeV/$c^2$ ($>$5 $\sigma$) from the nominal $K^0_S$ meson mass.
The $\pi^+\pi^-$ vertex is required to 
be displaced from the IP by a minimum distance of 0.22 cm.
The direction of the pion pair momentum in the transverse plane must agree with
the direction defined by the IP and the vertex displacement within 0.03 rad.
For $\pi^{0} \to \gamma\gamma$, a minimum photon energy of 50 MeV
is required and the $\gamma\gamma$ invariant mass must be within $\pm$16 MeV/$c^2$ ($\sim$2.5 $\sigma$)
of the nominal $\pi^{0}$ mass.

The decays $B_{\mathrm{sig}} \rightarrow K^+ \nu \overline{\nu}$, $\pi^+\nu\overline{\nu}$, $K_S^0 \nu\overline{\nu}$, and 
$\pi^0 \nu\overline{\nu}$ are reconstructed from single $K^+$, $\pi^+$, $K_S^0$, and $\pi^0$ candidates, respectively.
The $B^0 \rightarrow K^{*0}\nu\overline{\nu}$ candidate is reconstructed from a charged pion and 
an oppositely charged kaon,
while $B^+ \rightarrow K^{*+}\nu\overline{\nu}$ decays are reconstructed from a $K_S^0$ candidate and a charged pion, or
a charged kaon and a $\pi^0$ candidate. 
The reconstructed mass of the $K^{*0}$ ($K^{*+}$) candidate should be within a $\pm$75 MeV$/c^{2}$ window around
the nominal $K^{*0}$ ($K^{*+}$) mass. Furthermore, pairs of charged pions with opposite charge are used to form 
$B^0 \rightarrow \rho^0 \nu\overline{\nu}$ candidates where the $\pi^+\pi^-$ invariant mass should be 
within $\pm$150 MeV/$c^2$ from the nominal $\rho^0$ mass. For $B^+ \rightarrow \rho^+\nu\overline{\nu}$, 
a charged pion and a $\pi^0$ candidate are used, and a $\pm$150 MeV/$c^2$ mass window is required.
A $\phi$ meson is formed from a $K^+K^-$ pair with a 
reconstructed mass within $\pm$10 MeV/$c^2$ ($\sim$2 $\sigma$) from the nominal $\phi$ mass.

No additional charged tracks or $\pi ^{0}$
candidates are allowed in the event. 
We select $B_{\mathrm{sig}}$ candidates using
the variable $E_{\mathrm{ECL}}\equiv E_{\mathrm{tot}}-E_{\mathrm{rec}}$,
where $E_{\mathrm{tot}}$ and $E_{\mathrm{rec}}$ are the total visible energy
measured by the ECL detector and the measured energy of reconstructed objects
including the $B_{\mathrm{tag}}$ and the signal side $h^{(*)}$ candidate,
respectively. A minimum threshold of 50 (100, 150) MeV on the cluster energy
is applied for the barrel (forward endcap, backward endcap) region of the
ECL detector. 
The decays $B \to D^{*} \ell \nu$ are examined as control samples;
the observed $E_{\mathrm{ECL}}$ distributions are found to be in good agreement with Monte Carlo (MC) simulations~\cite{Ikado:2006un}.
The signal region is defined by $E_{\mathrm{ECL}}$~$<$~0.3~GeV
while the sideband region is given by 0.45~GeV~$<E_{\mathrm{ECL}}<$~1.5~GeV.

The dominant background source is 
$B\overline{B}$ decays involving a $b \to c$ transition.
A lower bound of 1.6 GeV/$c$
on $P^*$, the momentum of the $h^{(*)}$ candidate in the $B_{\mathrm{sig}}$ rest frame,
suppresses this background, while an upper bound of 2.5 GeV/$c$ rejects
the contributions from radiative two-body modes such as $B \to K^*\gamma$.
The $P^*$ requirement is removed for $\phi$ candidates due to the lack of 
theoretical calculations for $B_d \to \phi$ form factors.
Furthermore, the cosine of the angle between the missing momentum in the
laboratory frame and the beam is required to 
lie between $-0.86$ and $0.95$.
The missing momentum is calculated 
using the momenta of the reconstructed $B_{\mathrm{tag}}$ and $h^{(*)}$
candidates. These criteria suppress backgrounds with particles
produced along the beam pipe.
Other background sources, such as $e^+e^- \to q\overline{q}$~$(q=u,d,c,s)$ continuum background and 
rare $B$ decays involving $b \to u$, $b \to s$, or $b \to d$
processes, are found to be small. 

\begin{figure*}[htpb]
\begin{center}
\unitlength=1cm
\begin{picture}(14.4,3.2)
\put( 0.0,0.0){\includegraphics[width=4.8cm,height=3.2cm]{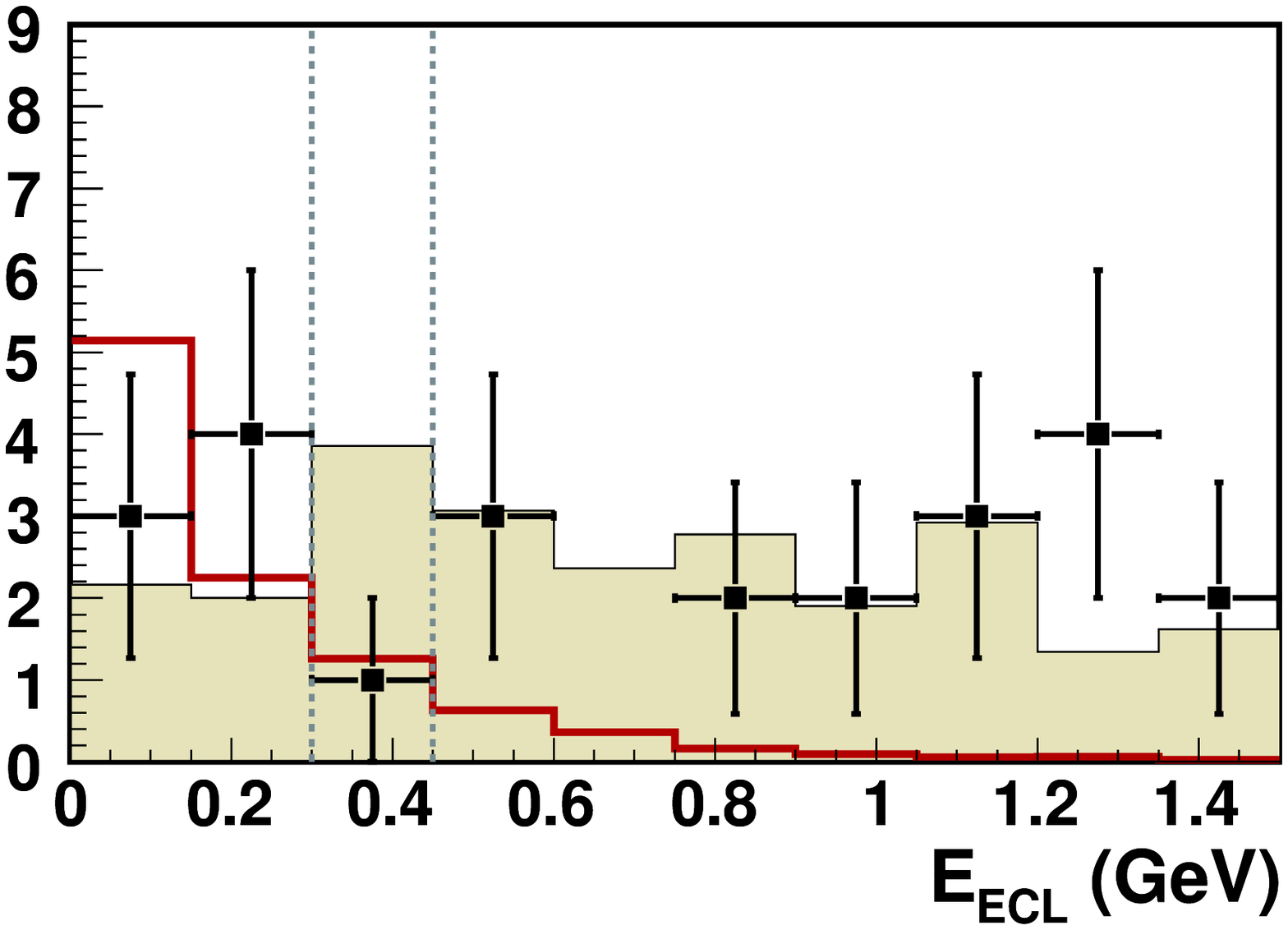}}
\put( 4.8,0.0){\includegraphics[width=4.8cm,height=3.2cm]{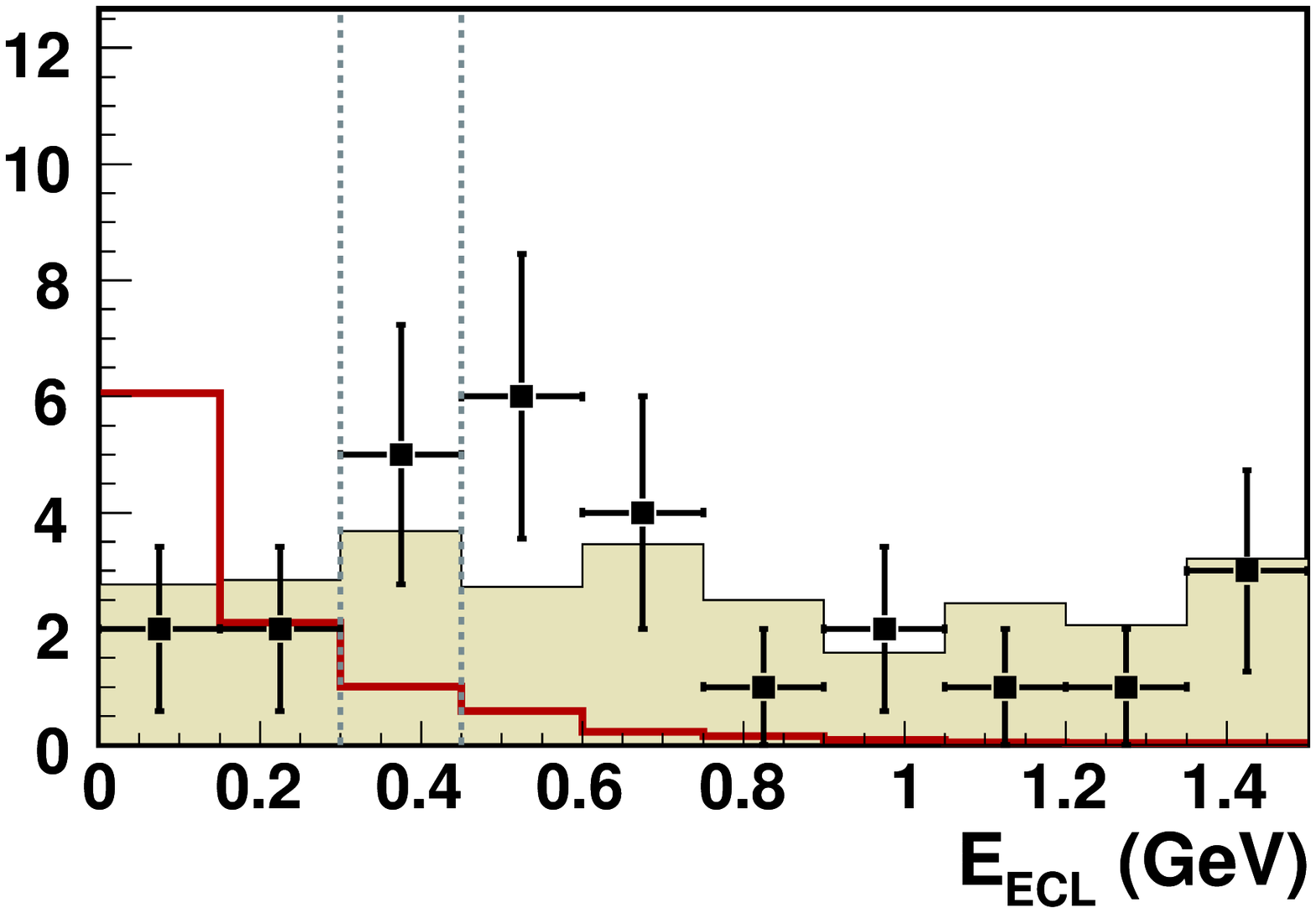}}
\put( 9.6,0.0){\includegraphics[width=4.8cm,height=3.2cm]{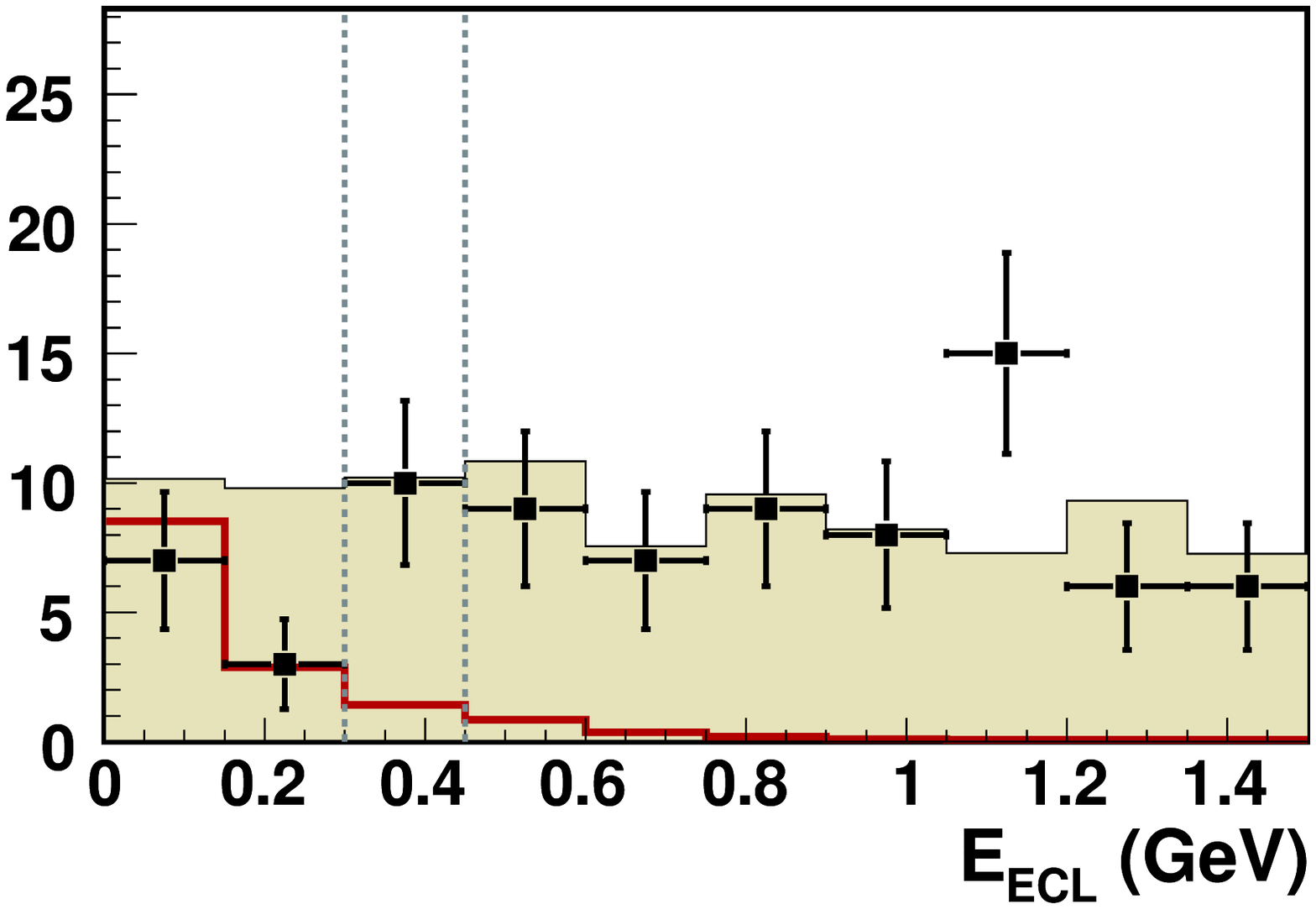}}
\put( 2.0,2.7){\footnotesize a) $B^0 \to K^{*0} \nu \overline{\nu}$}
\put( 6.8,2.7){\footnotesize b) $B^+ \to K^{*+} \nu \overline{\nu}$}
\put(11.6,2.7){\footnotesize c) $B^+ \to K^+  \nu \overline{\nu}$}
\end{picture}
\begin{picture}(14.4,3.2)
\put( 0.0,0.0){\includegraphics[width=4.8cm,height=3.2cm]{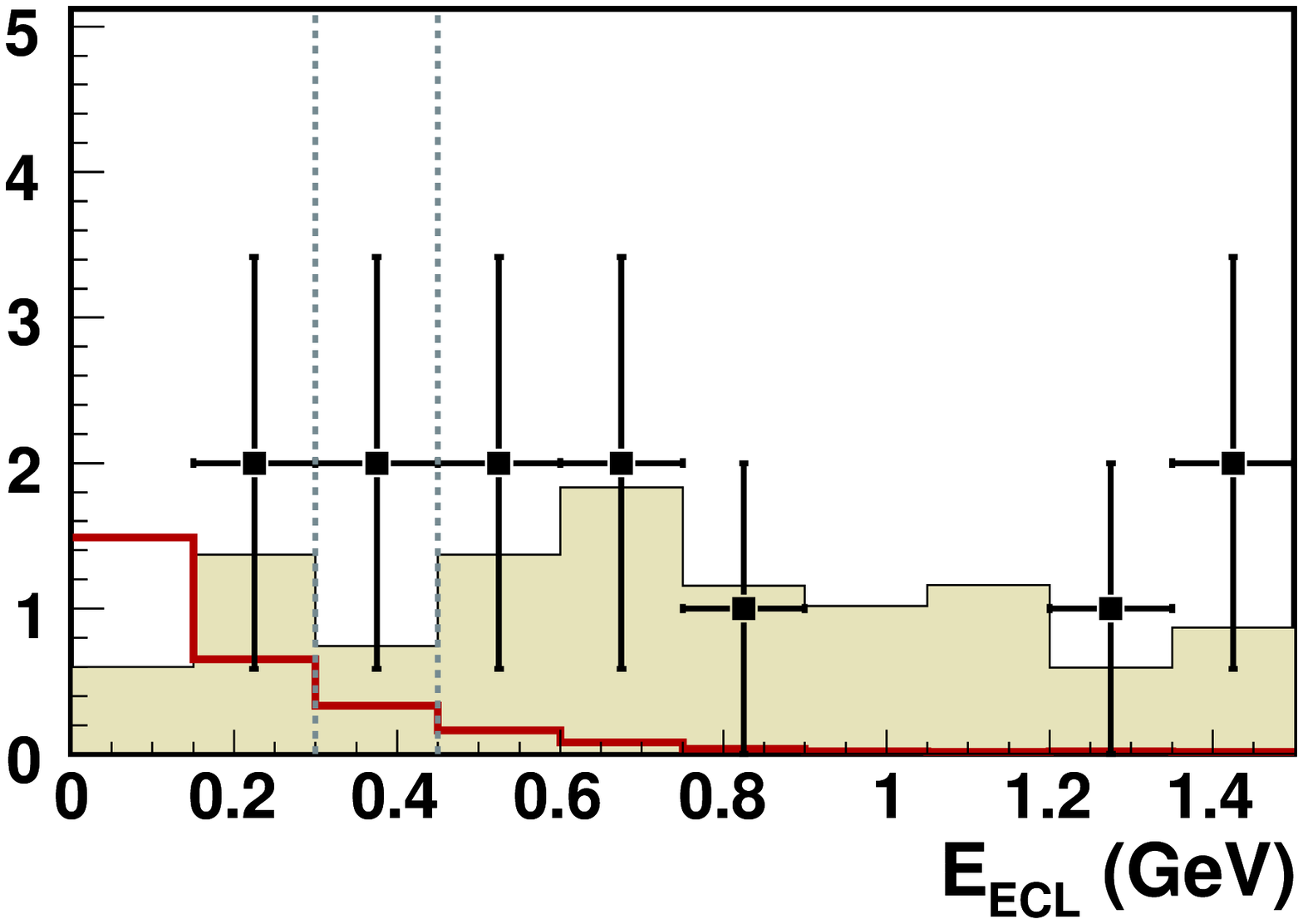}}
\put( 4.8,0.0){\includegraphics[width=4.8cm,height=3.2cm]{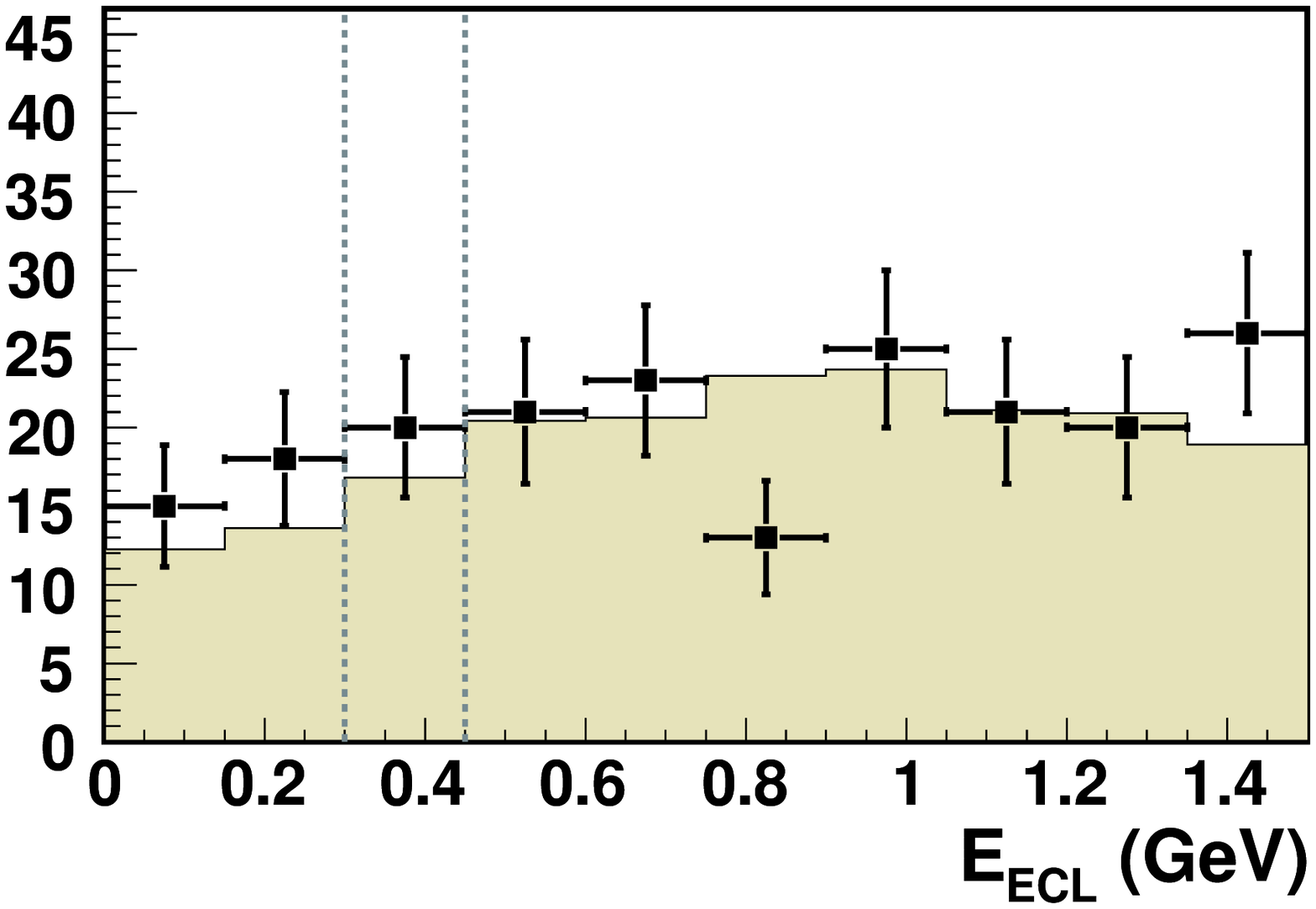}}
\put( 9.6,0.0){\includegraphics[width=4.8cm,height=3.2cm]{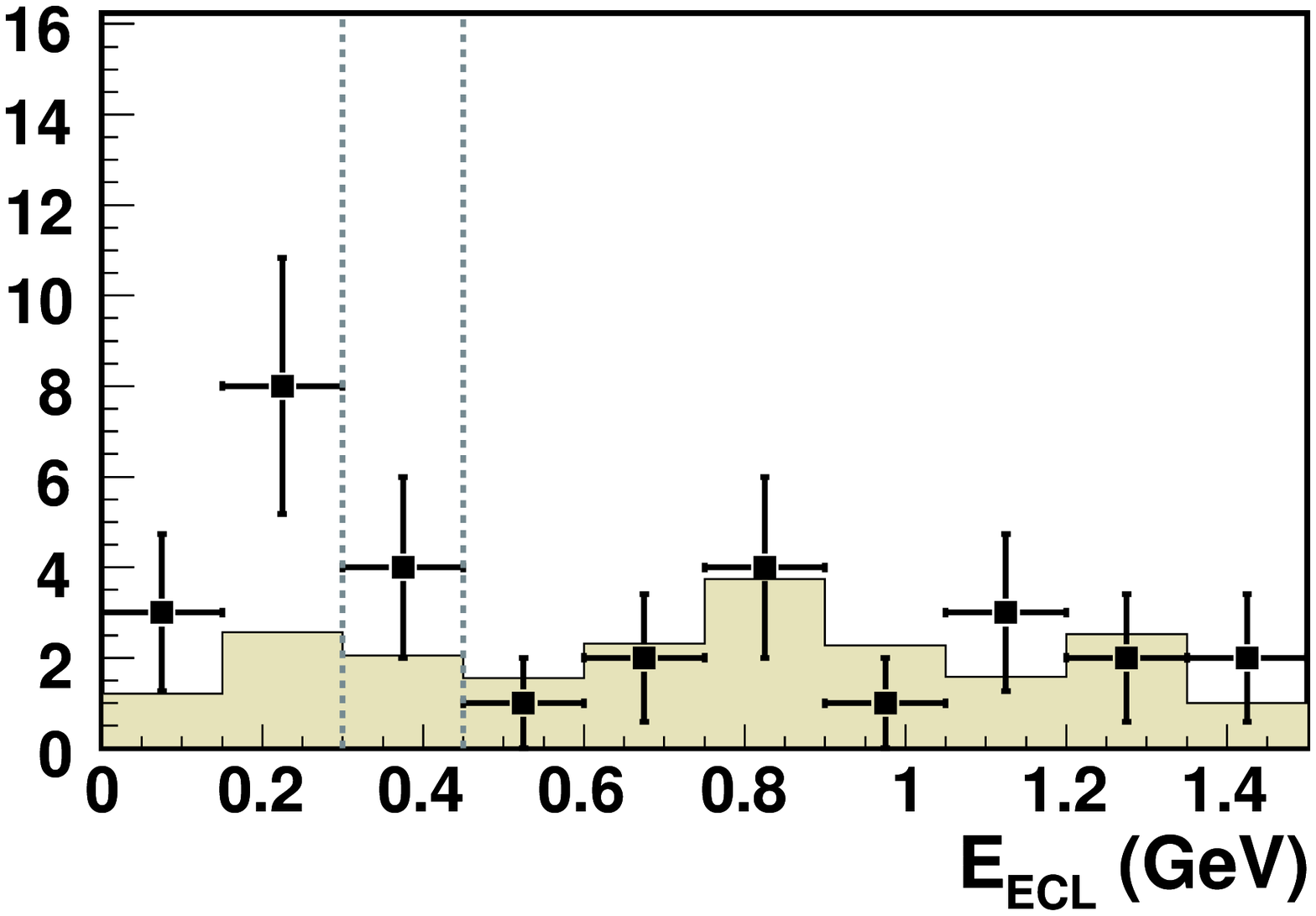}}
\put( 2.0,2.7){\footnotesize d) $B^0 \to K_S^0\nu \overline{\nu}$}
\put( 6.8,2.7){\footnotesize e) $B^+ \to \pi^+ \nu \overline{\nu}$}
\put(11.6,2.7){\footnotesize f) $B^0 \to \pi^0 \nu \overline{\nu}$}
\end{picture}
\begin{picture}(14.4,3.2)
\put( 0.0,0.0){\includegraphics[width=4.8cm,height=3.2cm]{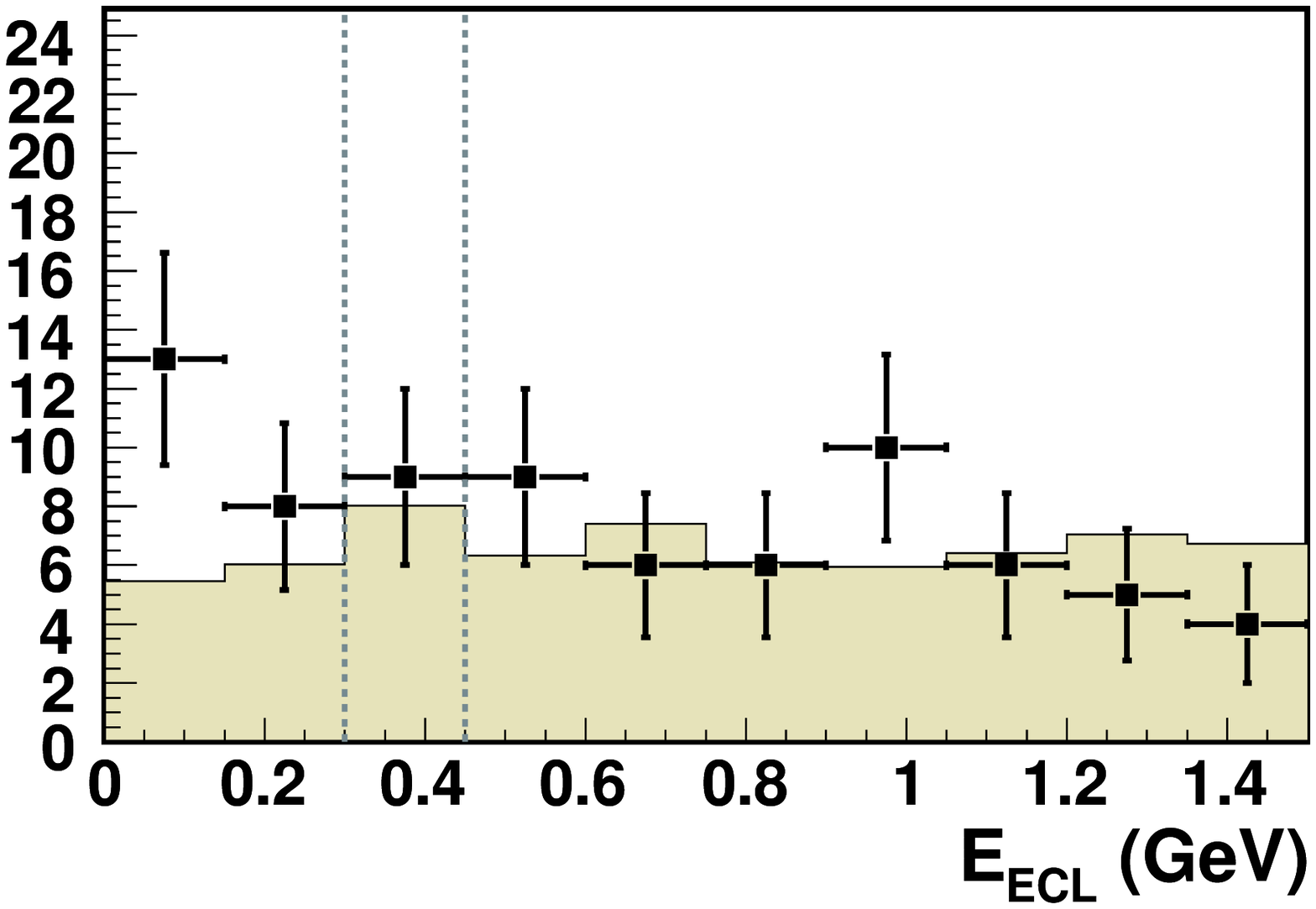}}
\put( 4.8,0.0){\includegraphics[width=4.8cm,height=3.2cm]{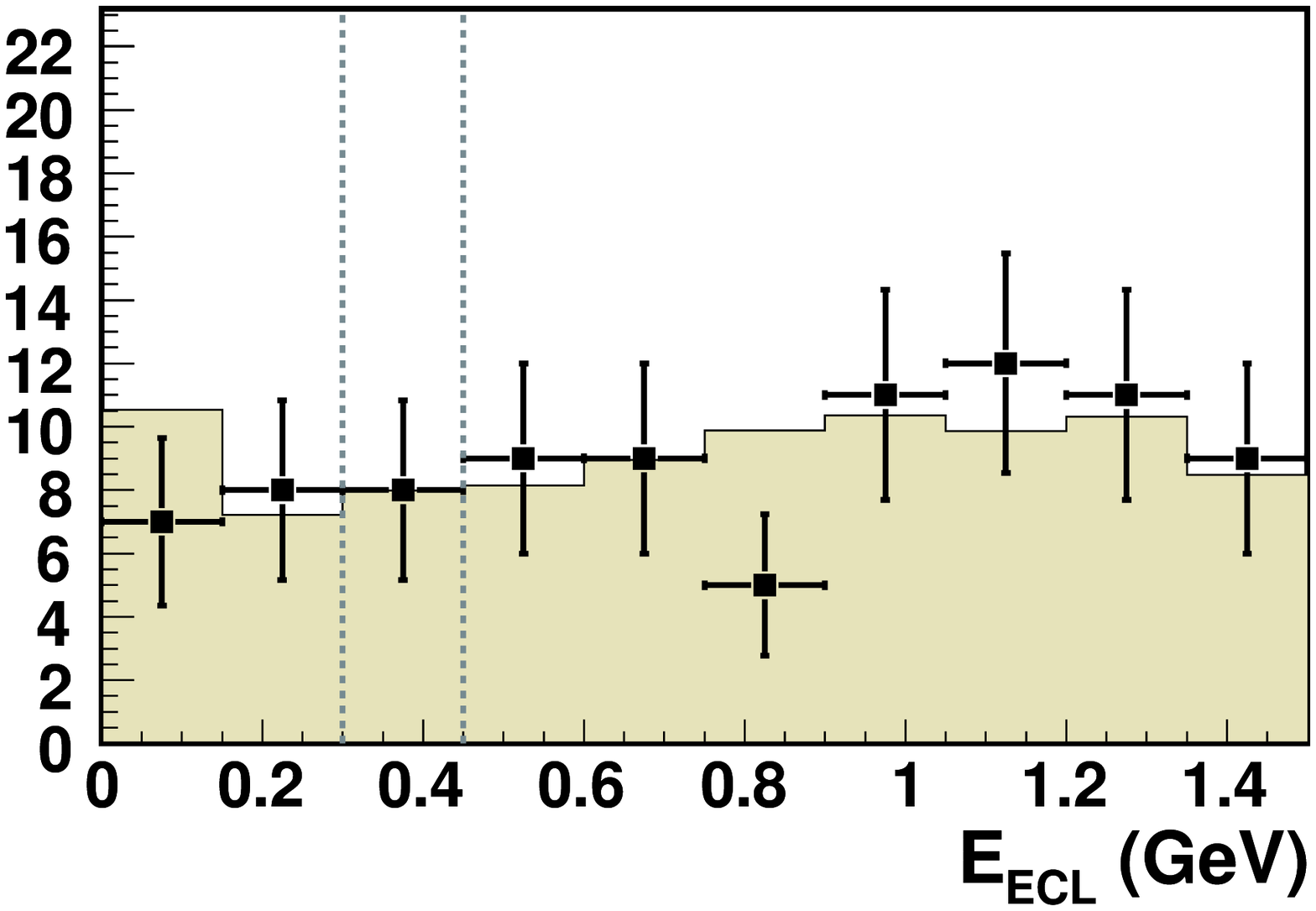}}
\put( 9.6,0.0){\includegraphics[width=4.8cm,height=3.2cm]{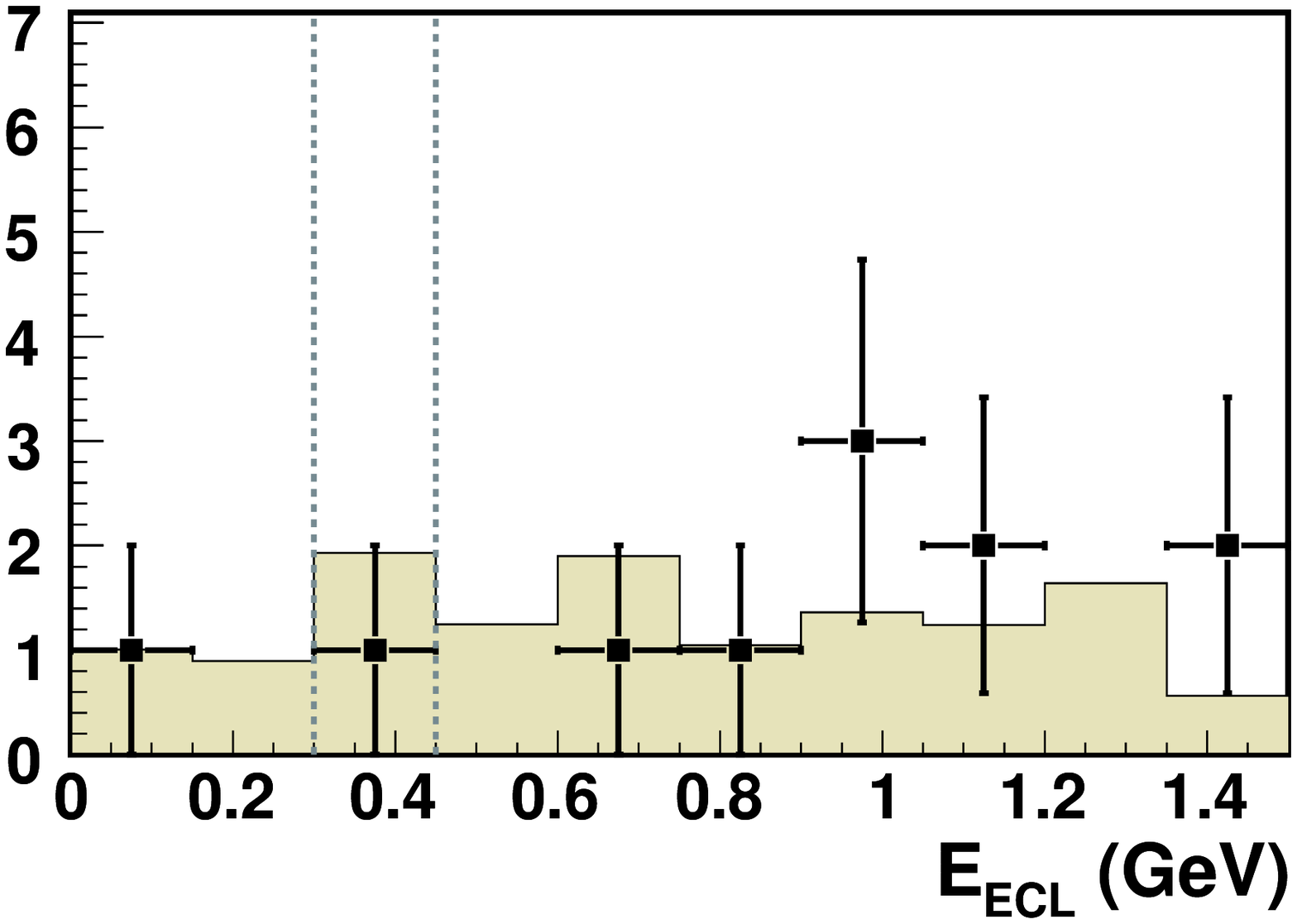}}
\put( 2.0,2.7){\footnotesize g) $B^0 \to \rho^0 \nu \overline{\nu}$}
\put( 6.8,2.7){\footnotesize h) $B^+ \to \rho^+ \nu \overline{\nu}$}
\put(11.6,2.7){\footnotesize i) $B^0 \to \phi \nu \overline{\nu}$}
\end{picture}
\end{center}
\caption{The $E_{\mathrm{ECL}}$ distributions for $B \to h^{(*)}\nu\overline{\nu}$ decays.
The shaded histograms show the background distributions from MC
simulations and are normalized to sideband data. 
The open histograms show the SM expected signal distributions for
$B \to K^{(*)}\nu\overline{\nu}$ decays multiplied by a factor of 20 for the comparison.
The vertical dashed lines show the upper bound (left) 
of the signal box and the lower bound (right) of the sideband region.}
\label{fig:final_plots}
\end{figure*}

The reconstructed $E_{\mathrm{ECL}}$ distributions are shown in Fig.~\ref{fig:final_plots}.
The $E_{\rm ECL}$ distributions of background are estimated with 
MC simulations; in particular, a large $b \to c$ MC sample corresponding to ten times the data luminosity 
is introduced with a preselection on the generator information.  The background $E_{\rm ECL}$ distributions 
are normalized by the number of events in the sideband region.
None of the signal modes has a significant signal. Including the effects of both statistical and
systematic uncertainties, an extension of the Feldman-Cousins 
method~\cite{Feldman:1997qc,Conrad:2002kn} is used to calculate the upper limits on the branching fractions.
The observed number of events in the signal box and sideband region, expected background 
contributions in the signal box, reconstruction efficiencies, and the obtained upper limits 
at 90\% confidence level (CL) are shown in Table~\ref{tab:yield_table}.
The reconstruction efficiencies are estimated with MC simulations using the $B \to h^{(*)}$ form factors from 
Ref.~\cite{ref:form_factors}. The $B^0 \to \phi\nu\overline{\nu}$ MC samples are generated with the $B \to K^{*}$ form factors.

\begin{table}[htpb]
\caption{A summary of the number of observed events in the signal box ($N_{\rm obs}$) and sideband regions ($N_{\rm side}$), 
expected background yields ($N_{b}$) in the signal box, reconstruction efficiencies including both $B_{\rm tag}$ and $B_{\rm sig}$ ($\epsilon$), 
and the upper limits (U.L.) on the branching fractions at 90\% CL.}
\label{tab:yield_table}
\begin{center}
\begin{tabular}{lrrrrr}
\hline
Mode~ & $N_{\rm obs}$ & $N_{\rm side}$ & $N_{b}$~~ & $\epsilon (\times 10^{-5})$ & U.L.~~ \\ 
\hline
$K^{*0}\nu\overline{\nu}$ &  7 &  16 &  $4.2\pm1.4$ &  $5.1\pm0.3$ & ~$<3.4\times 10^{-4}$ \\
$K^{*+}\nu\overline{\nu}$ &  4 &  18 &  $5.6\pm1.8$ &  $5.8\pm0.7$ & ~$<1.4\times 10^{-4}$ \\
$~~~\to K_S^0\pi^+$       &  1 &   7 &  $2.3\pm1.2$ &  $2.8\pm0.3$ &  \\
$~~~\to K^+\pi^0$         &  3 &  11 &  $3.3\pm1.4$ &  $3.0\pm0.4$ &  \\
$K^{+}\nu\overline{\nu}$  & 10 &  60 & $20.0\pm4.0$ & $26.7\pm2.9$ & ~$<1.4\times 10^{-5}$ \\
$K^0\nu\overline{\nu}$    &  2 &   8 &  $2.0\pm0.9$ &  $5.0\pm0.3$ & ~$<1.6\times 10^{-4}$ \\
$\pi^+\nu\overline{\nu}$  & 33 & 149 & $25.9\pm3.9$ & $24.2\pm2.6$ & ~$<1.7\times 10^{-4}$ \\
$\pi^0\nu\overline{\nu}$  & 11 &  15 &  $3.8\pm1.3$ & $12.8\pm0.8$ & ~$<2.2\times 10^{-4}$ \\
$\rho^0\nu\overline{\nu}$ & 21 &  46 & $11.5\pm2.3$ &  $8.4\pm0.5$ & ~$<4.4\times 10^{-4}$ \\
$\rho^+\nu\overline{\nu}$ & 15 &  66 & $17.8\pm3.2$ &  $8.5\pm1.1$ & ~$<1.5\times 10^{-4}$ \\
$\phi\nu\overline{\nu}$   &  1 &   9 &  $1.9\pm0.9$ &  $9.6\pm1.4$ & ~$<5.8\times 10^{-5}$ \\
\hline
\end{tabular}
\end{center}
\end{table} 

The uncertainties associated with the background yields are dominated by the data sideband statistics and MC statistics,
and are summarized in Table~\ref{tab:uncertainties1}. The possible disagreement
in the $E_{\rm ECL}$ distributions
between data and MC is checked using 
wrong-flavor combinatorial events, and an uncertainty of 0.1--2.0 events is included. 
Possible backgrounds from rare $B$ decays are examined using a large MC sample corresponding to 
50 times the data luminosity. We change the relative normalizations 
of rare $B$ components by $\pm$50\%, and 
the variation in the background yield (0.1--1.8 events) is included as a systematic uncertainty.

Various sources of uncertainties are considered for the signal normalization and are summarized in 
Table~\ref{tab:uncertainties2}.
The uncertainties in $B_{\mathrm{tag}}$ reconstruction (2.0\% for $B^0$ and 9.9\% for $B^\pm$) 
are estimated by comparing the yields of data and MC from the $B_{\mathrm{tag}}$ candidates.
Systematic uncertainty arising from the track and $\pi^0$ rejection is studied using $B \to D^{(*)}\ell\nu$
decays, and an error of 2.7\% is assigned. 
The uncertainties in the efficiencies for detecting a $K_S^0$ or $\pi^0$ from $B_{\mathrm{sig}}$
are estimated to be 4.9\% and 4.0\%, respectively.
We also vary the $B \to h^{(*)}$ form factors used in the signal MC generation according to the uncertainties
given by the Ref.~\cite{ref:form_factors}, and an uncertainty of 0.4--3.7\% on the reconstruction efficiency is included.
A larger uncertainty of 13\%, which is estimated from the difference between the default decay model and a generic three-body
phase-space model, is introduced for $B^0 \to \phi \nu\overline{\nu}$ decays. 
Furthermore, the following uncertainties are also considered: 
the number of $B\overline{B}$ events (1.3\%), tracking efficiency (1.0--2.2\%), 
particle identification (0.5--2.0\%), $h^{(*)}$ mass selection (0.8--2.3\%), and
the $\phi\to K^+K^-$ branching fraction (1.2\%).

\begin{table*}[htpb]
\caption{Summary of the uncertainties associated with the background yields (in the number of events).}
\label{tab:uncertainties1}
\begin{center}
\begin{tabular}{lccccccccc}
\hline
Uncertainties & 
$K^{*0}\nu\overline{\nu}$~ & 
$K^{*+}\nu\overline{\nu}$~ & 
$K^{+}\nu\overline{\nu}$~  & 
$K^0\nu\overline{\nu}$~    & 
$\pi^+\nu\overline{\nu}$~  & 
$\pi^0\nu\overline{\nu}$~  & 
$\rho^0\nu\overline{\nu}$~ & 
$\rho^+\nu\overline{\nu}$~ & 
$\phi\nu\overline{\nu}$    \\
\hline
Sideband statistics & 1.0 & 1.3 & 2.6 & 0.7 & 2.1 & 1.0 & 1.7 & 2.2 & 0.6 \\
MC statistics	    & 0.9 & 1.2 & 2.6 & 0.6 & 2.0 & 0.8 & 1.3 & 1.6 & 0.6 \\
MC/data difference  & 0.3 & 0.3 & 1.5 & 0.2 & 2.0 & 0.3 & 0.9 & 1.4 & 0.1 \\
Rare $B$	    & 0.1 & 0.3 & 0.2 & 0.2 & 1.8 & 0.1 & 0.1 & 0.9 & 0.3 \\
\hline
Total		    & 1.4 & 1.8 & 4.0 & 0.9 & 3.9 & 1.3 & 2.3 & 3.2 & 0.9 \\
\hline
\end{tabular}
\end{center}
\end{table*} 

\begin{table*}[htpb]
\caption{Summary of the relative uncertainties for signal normalization (in \%).}
\label{tab:uncertainties2}
\begin{center}
\begin{tabular}{lccccccccc}
\hline
Source & 
$K^{*0}\nu\overline{\nu}$~ & 
$K^{*+}\nu\overline{\nu}$~ & 
$K^{+}\nu\overline{\nu}$~  & 
$K^0\nu\overline{\nu}$~    & 
$\pi^+\nu\overline{\nu}$~  & 
$\pi^0\nu\overline{\nu}$~  & 
$\rho^0\nu\overline{\nu}$~ & 
$\rho^+\nu\overline{\nu}$~ & 
$\phi\nu\overline{\nu}$    \\
\hline
$N(B\overline{B})$ & 
1.3 & 1.3 & 1.3 & 1.3 & 1.3 & 1.3 & 1.3 & 1.3 & 1.3 \\
Tracking efficiency & 
2.1 & 1.1 & 1.0 & -   & 1.0 & -   & 2.2 & 1.1 & 2.0 \\
$K_S^0$/$\pi^0$ reconstruction & 
-   & 4.4 & -   & 4.9 & -   & 4.0 & -   & 4.0 & -   \\
Sub branching fraction & 
-   & -   & -	& -   & -   & -   & -	& -   & 1.2 \\
Particle identification &
1.3 & 0.7 & 0.7 & -   & 0.5 & -   & 1.0 & 0.5 & 2.0 \\
MC statistics &
3.5 & 2.4 & 1.9 & 3.2 & 2.0 & 2.8 & 3.3 & 3.4 & 2.3 \\
Mass selection &
0.8 & 2.3 & -   & -   & -   & -   & 1.1 & 2.6 & 2.0 \\
$B_{\mathrm{tag}}$ reconstruction &
2.0 & 9.9 & 9.9 & 2.0 & 9.9 & 2.0 & 2.0 & 9.9 & 2.0 \\
Track/$\pi^0$ rejection &
2.7 & 2.7 & 2.7 & 2.7 & 2.7 & 2.7 & 2.7 & 2.7 & 2.7 \\
$B \to h^{(*)}$ form factor &
1.6 & 1.3 & 2.6 & 0.4 & 3.0 & 1.0 & 1.7 & 3.7 & 13.0 \\
\hline
Total &
5.9 & 11.8 & 10.9 & 6.9 & 11.0 & 6.1 & 5.8 & 12.5 & 14.2 \\
\hline
\end{tabular}
\end{center}
\end{table*} 

In conclusion, we have performed a search for  
$B \to h^{(*)} \nu \overline{\nu}$ decays with a fully reconstructed $B$ tagging method on a
data sample of $535\times 10^{6}$ $B\bar{B}$ pairs collected at the 
$\Upsilon(4S)$ resonance with the Belle detector. 
No significant signal is observed and we set upper limits on the branching fractions
at 90\% CL. The limits obtained for $B^0 \to K^{*0}\nu\overline{\nu}$ and $B^+ \to K^+\nu\overline{\nu}$ decays
are more stringent than the previous constraints from DELPHI~\cite{Adam:1996ts} and BaBar~\cite{Aubert:2004ws}.
The first searches for
$B^0 \to K^0 \nu\overline{\nu}$, $\pi^0\nu\overline{\nu}$, $\rho^0\nu\overline{\nu}$, $\phi\nu\overline{\nu}$, and
$B^+ \to K^{*+}\nu\overline{\nu}$, $\rho^+\nu\overline{\nu}$ are carried out, and upper limits
on the branching fraction of order $10^{-4}$ are obtained.
The limit on $B^+ \to \pi^+\nu\overline{\nu}$ is less restrictive than BaBar's result~\cite{Aubert:2004ws} due to 
a larger number of observed events in the signal box.
The results on $B \to K^{(*)} \nu \overline{\nu}$ reported
here are one order of magnitude above the predictions of Buchalla {\it et al.}~\cite{ref:buchalla} and 
hence still allow room for substantial non-SM contributions.
A higher luminosity $B$-factory experiment is required to probe the SM predictions for the branching fractions.

We thank the KEKB group for excellent operation of the
accelerator, the KEK cryogenics group for efficient solenoid
operations, and the KEK computer group and
the NII for valuable computing and Super-SINET network
support.  We acknowledge support from MEXT and JSPS (Japan);
ARC and DEST (Australia); NSFC and KIP of CAS (China); 
DST (India); MOEHRD, KOSEF, KRF and SBS Foundation (Korea); 
KBN (Poland); MES and RFAAE (Russia); ARRS (Slovenia); SNSF (Switzerland); 
NSC and MOE (Taiwan); and DOE (USA).


\end{document}